\documentclass[lettersize, journal]{IEEEtran}
\usepackage{amsmath,amsfonts}
\usepackage{array}
\usepackage[caption=false,font=normalsize,labelfont=sf,textfont=sf]{subfig}
\usepackage{textcomp}
\usepackage{stfloats}
\usepackage{url}
\usepackage{verbatim}
\usepackage{graphicx}
\usepackage{cite}
\usepackage{amsmath,amssymb}
\usepackage{algorithm}
\usepackage{algorithmic}
\usepackage{subfig} 
\usepackage{makecell}
\usepackage[top=0.56in, bottom=0.74in, left=0.48in, right=0.48in]{geometry}
\usepackage{multirow} 

 \usepackage{tabularx}
 \usepackage{array}

\usepackage{amsthm}

\usepackage{graphicx} 
\usepackage{amsmath}
\usepackage{amsmath}
\usepackage{amsmath, amssymb}
\usepackage{xcolor}	
\usepackage{amsmath} 

\usepackage{graphics}
\usepackage{amsmath, amsthm}

\theoremstyle{plain}

\begin{document}

\title{Adaptive DOA Estimation With KL-Based Kernel Bandwidth Adaptation Under Impulsive Noise}

\author{
Mojtaba Hajiabadi, Saeid Pakravan, Ghosheh Abed Hodtani, Ming Zeng

\thanks{
M. Hajiabadi is with the Department of Electrical \& Computer Engineering, University of Birjand, Birjand, Iran.
email: mhajiabadi@birjand.ac.ir.
}

\thanks{
S. Pakravan is with the Department of Computer Science, University of Quebec in Montreal (UQAM), Montreal, QC, Canada.
email: pakravan.saeid@uqam.ca.
}

\thanks{G. Abed Hodtani is with the Department of Electric and Computer Engineering, Ferdowsi University, Mashhad, Iran. email: hodtani@um.ac.ir.}

 \thanks{M. Zeng is with the Department of Electric and Computer Engineering, Laval University, Quebec City, QC, Canada. email: ming.zeng@gel.ulaval.ca.}

}

	\maketitle

\begin{abstract}
Direction-of-arrival (DOA) estimation can degrade markedly in
impulsive noise, where methods relying on Gaussian or second-order
statistical assumptions are particularly susceptible to
large-amplitude outliers. This paper develops an adaptive DOA
estimator that employs Gaussian kernel weighting to attenuate the
influence of large prediction errors while adjusting the kernel
bandwidth online according to a Kullback--Leibler (KL) divergence
criterion. A sample-based approximation of the KL criterion yields a
recursive bandwidth update that adapts the degree of outlier
suppression to the observed error statistics rather than prescribing
a fixed bandwidth a priori. The spatial-filter coefficients and
kernel bandwidth are updated recursively, and the DOAs are recovered
from the resulting spatial spectrum. Mean and mean-square analyses
are developed under standard adaptive-filtering approximations to
characterize the coefficient convergence and steady-state behavior.
Simulation results under Gaussian-mixture impulsive noise demonstrate
stable coefficient adaptation and improved DOA estimation accuracy
over representative competing methods in several challenging
operating regimes, with the most pronounced gains occurring at
low-to-moderate signal-to-noise ratios. The results also indicate
reduced sensitivity to the initial kernel bandwidth through online
bandwidth adaptation.

\end{abstract}

\begin{IEEEkeywords}
Direction-of-arrival (DOA) estimation, impulsive noise,
Kullback--Leibler (KL) divergence, adaptive kernel bandwidth,
robust adaptive filtering.
\end{IEEEkeywords}

\section{Introduction}
\label{sec:introduction}

Direction-of-arrival (DOA) estimation is a fundamental task in array
signal processing, with important applications in wireless
communications, array-based localization, acoustic sensing, and
related signal-processing systems. It also remains an active research
topic in modern multiple-input multiple-output (MIMO) communication
systems, where practical hardware constraints can affect estimation
performance \cite{11052726}. Classical high-resolution techniques,
such as multiple signal classification (MUSIC) and estimation of
signal parameters via rotational invariance techniques (ESPRIT), can
achieve high angular resolution under conventional modeling
assumptions \cite{Schmidt1986,Roy1989}. However, conventional
subspace implementations rely on sample covariance or related
second-order statistical information and can therefore be sensitive
to impulsive disturbances \cite{Guo2023VBI}. In heavy-tailed
environments, a small fraction of large-amplitude observations can
substantially distort sample covariance estimates; moreover,
second-order moments may not exist for some heavy-tailed models.
Consequently, DOA estimators that rely on second-order statistical
information can experience substantial performance degradation under
impulsive noise.

Robust DOA estimation under non-Gaussian and impulsive noise has
therefore received considerable attention. Variational Bayesian
approaches have been developed to mitigate the effects of outliers
and heavy-tailed disturbances \cite{Liu2022RobustVBI,Guo2023VBI},
while other methods exploit fractional lower-order statistics,
robust nonlinear criteria, or characteristic-function-based
representations to reduce sensitivity to large-amplitude observations
\cite{Li2018FLOS,Zhang2017Cauchy,Asghari2022ECF,Wang2023RobustDOA}.
More recently, learning-based architectures have been explored for
robust DOA estimation under challenging operating conditions
\cite{10878805}. Adaptive approaches have also received attention,
including robust criteria with data-dependent tuning parameters
\cite{Dai2025ASE} and adaptive-filter-based DOA estimators designed
to improve online coefficient adaptation and mitigate noise-induced
estimation bias \cite{RecentBCNLMS2024}. More recently, \cite{11535307} investigated variable step-size general bias-compensated M-estimate adaptive filtering for robust DOA estimation, highlighting a complementary direction in robust adaptive array processing. Collectively, these studies
demonstrate the benefits of robust and adaptive processing while
highlighting that estimator performance can depend strongly on the
adopted criterion, estimation architecture, and associated tuning
parameters.

Kernel-based information-theoretic criteria, particularly
correntropy and its generalizations, provide a robust alternative
for processing data contaminated by impulsive disturbances. By
assigning smaller kernel-induced weights to large residuals,
correntropy-based methods can reduce the influence of outlying
observations and have been applied to DOA estimation in impulsive
noise \cite{Zhang2014CorrentropyDOA,Ma2020ComplexCorrentropy,
Tian2019GCO, Alizadeh2025Channel}. Their performance, however, depends strongly on the
kernel bandwidth \cite{Tian2019GCO}. A small bandwidth provides
stronger attenuation of large residuals but may excessively
downweight informative samples, whereas a large bandwidth weakens
outlier suppression and approaches quadratic-error behavior.
Consequently, a fixed bandwidth imposes a tradeoff among robustness,
convergence, and steady-state accuracy and may be inadequate when the
error statistics vary across operating conditions or during
adaptation.

Adaptive kernel-bandwidth mechanisms have been investigated to
alleviate the limitations of fixed-bandwidth processing. In DOA
estimation, a data-dependent kernel-size function has been incorporated
into generalized-autocorrentropy-based processing
\cite{Tian2019GCO}, while more recent approaches have employed
adaptive kernel-width mixture correntropy in subspace-based
\cite{Dai2025TGRS} and adaptive nulling-array formulations
\cite{Sun2025JASA}. In a broader information-theoretic learning
context, Singh and Pr\'{\i}ncipe developed an online kernel-width
adaptation rule by minimizing a Kullback--Leibler (KL) divergence
criterion involving the error density and its kernel estimate
\cite{SinghPrincipe2011}. These studies establish adaptive kernel
bandwidth selection in both DOA estimation and adaptive
information-theoretic learning, but employ different bandwidth
criteria and estimation architectures. 

Motivated by this distinction, we develop an online kernel-weighted
DOA estimator that combines reference-sensor spatial-filter
adaptation with KL-based kernel-bandwidth adaptation. The
reference-sensor prediction error drives the spatial-filter update,
while Gaussian kernel weighting attenuates the influence of large
residuals. Rather than prescribing the kernel bandwidth a priori, a
sample-based approximation of the KL criterion is used to update it
recursively from recent prediction errors. The spatial-filter
coefficients and kernel bandwidth are therefore updated within the
same adaptive procedure, allowing the error-dependent weighting to
evolve during adaptation. The DOAs are subsequently recovered from
the spatial spectrum constructed using the adapted spatial filter.

\begin{table*}[!t]
\centering
\caption{Comparison of Representative Robust DOA Estimation
Approaches Under Impulsive Noise}
\label{tab:comparison}
\renewcommand{\arraystretch}{1.18}
\setlength{\tabcolsep}{3.5pt}

\begin{tabular}{
    >{\centering\arraybackslash}p{1.25cm}
    >{\centering\arraybackslash}p{1.8cm}
    >{\centering\arraybackslash}p{2.0cm}
    >{\centering\arraybackslash}p{2.1cm}
    >{\raggedright\arraybackslash}p{10.8cm}
}
\hline
\textbf{Ref.}
&
\textbf{Adaptive Kernel BW}
&
\textbf{Online Spatial Adaptation}
&
\textbf{KL-Based BW Adaptation}
&
\textbf{Key Characteristic}
\\
\hline

\cite{11535307}
&
\texttimes
&
\checkmark
&
\texttimes
&
Variable step-size bias-compensated M-estimate
adaptive filtering for robust DOA estimation
\\
\cite{Zhang2014CorrentropyDOA}
&
\texttimes
&
\texttimes
&
\texttimes
&
Correntropy-based robust covariance estimation with MUSIC
\\

\cite{Ma2020ComplexCorrentropy}
&
\texttimes
&
\texttimes
&
\texttimes
&
Generalized complex correntropy with MUSIC
\\

\cite{Tian2019GCO}
&
\checkmark
&
\texttimes
&
\texttimes
&
Generalized autocorrentropy with adaptive kernel size
\\

\cite{Dai2025TGRS}
&
\checkmark
&
\texttimes
&
\texttimes
&
Adaptive kernel-width mixture correntropy with MUSIC
\\

\cite{Sun2025JASA}
&
\checkmark
&
\checkmark
&
\texttimes
&
Adaptive kernel-width mixture correntropy with adaptive nulling
\\

\hline
\textbf{This work}
&
\checkmark
&
\checkmark
&
\checkmark
&
KL-based bandwidth adaptation with online spatial filtering
\\
\hline

\end{tabular}
\end{table*}

The main contributions of this work are summarized as follows:
\begin{enumerate}
    \item An online kernel-weighted DOA estimator is developed by
    integrating reference-sensor spatial-filter adaptation with
    KL-based kernel-bandwidth adaptation, allowing the coefficient
    weighting and bandwidth to evolve from the observed prediction
    errors.

    \item A recursive bandwidth update is derived from a sample-based
    approximation of a KL-divergence criterion, thereby avoiding the
    need to prescribe a fixed kernel bandwidth throughout adaptation.

    \item Mean and mean-square analyses are developed under standard
    adaptive-filtering approximations, yielding a mean-stability
    condition and an approximate steady-state mean-square deviation
    (MSD) characterization.

    \item Numerical experiments under Gaussian-mixture impulsive noise
    assess the proposed estimator in terms of DOA accuracy, adaptation
    behavior, and parameter sensitivity against representative robust
    DOA methods.
\end{enumerate}

The remainder of this paper is organized as follows.
Section~II presents the array signal model and adaptive DOA
formulation. Section~III develops the kernel-weighted coefficient
and KL-based bandwidth updates. Section~IV analyzes the mean and
mean-square behavior of the proposed estimator and discusses its
computational complexity. Section~V presents the numerical results.
Finally, Section~VI concludes the paper.

\textbf{Notation:} Lowercase boldface letters denote vectors, whereas uppercase boldface letters denote matrices. The operators $(\cdot)^{T}$, $(\cdot)^{*}$, and $(\cdot)^{H}$ denote transpose, complex conjugation, and Hermitian transpose, respectively. The symbols $\mathbb{R}$ and $\mathbb{C}$ denote the sets of real and complex numbers, respectively, and $\mathbb{E}\{\cdot\}$ denotes statistical expectation. The notation $|\cdot|$ denotes the magnitude of a scalar, whereas $\|\cdot\|_2$ denotes the Euclidean norm of a vector. The identity matrix of size $M$ is denoted by $\mathbf{I}_M$.

\section{Signal Model and Problem Formulation}
\label{sec:signal_model}

\subsection{Array Signal Model}
\label{subsec:array_model}

Consider a uniform linear array (ULA) of $M$ sensors with interelement
spacing $d$, receiving $Q<M$ narrowband far-field signals from
distinct directions
$\boldsymbol{\Theta}=[\theta_1,\ldots,\theta_Q]^T$. The complex
baseband observation at the $m$th sensor and snapshot $n$ is
\begin{equation}
x_m(n)
=
\sum_{q=1}^{Q}
s_q(n)
\exp\!\left(
-j\frac{2\pi m d}{\lambda}\sin\theta_q
\right)
+v_m(n),
\label{eq:sensor_model}
\end{equation}
where $m=0,\ldots,M-1$, $s_q(n)$ is the $q$th source signal,
$\lambda$ denotes the wavelength, and $v_m(n)$ is the additive noise
at the $m$th sensor.

Collecting the sensor observations yields
\begin{equation}
\mathbf{x}_n
=
\mathbf{A}(\boldsymbol{\Theta})\mathbf{s}_n+\mathbf{v}_n,
\label{eq:array_model}
\end{equation}
where
$\mathbf{x}_n=[x_0(n),\ldots,x_{M-1}(n)]^T\in\mathbb{C}^{M}$,
$\mathbf{s}_n=[s_1(n),\ldots,s_Q(n)]^T\in\mathbb{C}^{Q}$, and
$\mathbf{v}_n=[v_0(n),\ldots,v_{M-1}(n)]^T\in\mathbb{C}^{M}$.
The array manifold matrix is
\[
\mathbf{A}(\boldsymbol{\Theta})
=
[\mathbf{a}(\theta_1),\ldots,\mathbf{a}(\theta_Q)]
\in\mathbb{C}^{M\times Q},
\]
with steering vector
\begin{equation}
\mathbf{a}(\theta)
=
\left[
1,\,
e^{-j\omega(\theta)},\,
\ldots,\,
e^{-j(M-1)\omega(\theta)}
\right]^T,
\label{eq:steering_vector}
\end{equation}
where
$
\omega(\theta)
=
\frac{2\pi d}{\lambda}\sin\theta
\label{eq:spatial_frequency}
$
denotes the interelement spatial phase shift.

\subsection{Adaptive DOA Formulation}
\label{subsec:adaptive_doa}

\begin{figure}[t]
    \centering
    \includegraphics[width=8.5cm,height=5cm]{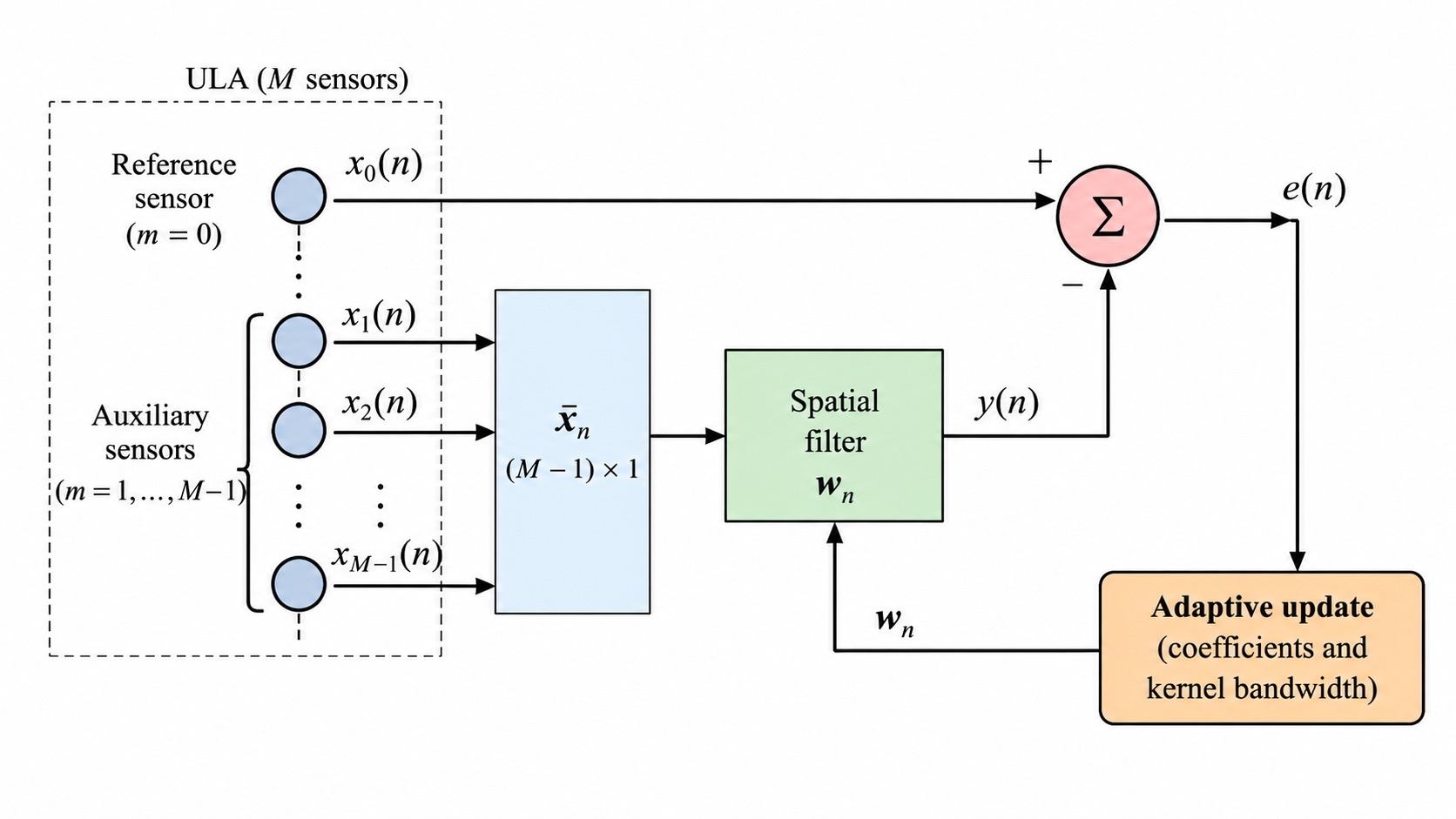}
    \caption{Reference-sensor adaptive array configuration for DOA estimation.}
    \label{fig:adaptive_array}
\end{figure}

To formulate DOA estimation as an adaptive prediction problem, the
sensor indexed by $m=0$ is selected as the reference sensor, while
the remaining $M-1$ sensors form the auxiliary subarray. Define
$
\bar{\mathbf{x}}_n
=
[x_1(n),\ldots,x_{M-1}(n)]^T
\in\mathbb{C}^{M-1}.
$
A complex-valued spatial filter with coefficient vector
$\mathbf{w}_n\in\mathbb{C}^{M-1}$ forms the prediction
$y(n)=\mathbf{w}_n^H\bar{\mathbf{x}}_n$ of the reference-sensor
observation. The corresponding prediction error is
\begin{equation}
e(n)
=
x_0(n)-\mathbf{w}_n^H\bar{\mathbf{x}}_n .
\label{eq:prediction_error}
\end{equation}
The reference-sensor prediction architecture is illustrated in
Fig.~\ref{fig:adaptive_array}.

To relate the prediction model to the source directions, the steering
vector in \eqref{eq:steering_vector} is partitioned as
\[
\mathbf{a}(\theta)
=
\begin{bmatrix}
1\\
\bar{\mathbf{a}}(\theta)
\end{bmatrix},
\qquad
\bar{\mathbf{a}}(\theta)
=
\left[
e^{-j\omega(\theta)},\ldots,
e^{-j(M-1)\omega(\theta)}
\right]^T .
\]
Accordingly, the array manifold matrix can be written as
\begin{equation}
\mathbf{A}(\boldsymbol{\Theta})
=
\begin{bmatrix}
\mathbf{1}_Q^T\\
\bar{\mathbf{A}}(\boldsymbol{\Theta})
\end{bmatrix},
\label{eq:partitioned_manifold}
\end{equation}
where
$\bar{\mathbf{A}}(\boldsymbol{\Theta})
=
[\bar{\mathbf{a}}(\theta_1),\ldots,
\bar{\mathbf{a}}(\theta_Q)]
\in\mathbb{C}^{(M-1)\times Q}$ and
$\mathbf{1}_Q$ is the $Q$-dimensional all-ones vector. It follows that
\[
x_0(n)
=
\mathbf{1}_Q^T\mathbf{s}_n+v_0(n),
\qquad
\bar{\mathbf{x}}_n
=
\bar{\mathbf{A}}(\boldsymbol{\Theta})\mathbf{s}_n
+\bar{\mathbf{v}}_n,
\]
where
$\bar{\mathbf{v}}_n=[v_1(n),\ldots,v_{M-1}(n)]^T$.

For a candidate direction $\theta$, the reference-sensor response is
unity, whereas the response predicted from the auxiliary subarray is
$\mathbf{w}_n^H\bar{\mathbf{a}}(\theta)$. Hence, directions satisfying
the adapted prediction relation produce small values of
$\left|1-\mathbf{w}_n^H\bar{\mathbf{a}}(\theta)\right|$. This
motivates the spatial spectrum
\begin{equation}
P_n(\theta)
=
\frac{1}
{\left|1-\mathbf{w}_n^H\bar{\mathbf{a}}(\theta)\right|^2}.
\label{eq:spatial_spectrum}
\end{equation}
After convergence, the DOAs are estimated from the dominant local
maxima of \eqref{eq:spatial_spectrum} over the prescribed angular
search region. When the number of sources $Q$ is known, the $Q$
largest distinct peaks are selected to obtain
$\widehat{\boldsymbol{\Theta}}
=[\hat{\theta}_1,\ldots,\hat{\theta}_Q]^T$.

The remaining problem is therefore to adapt $\mathbf{w}_n$ robustly
in the presence of impulsive noise. The next section develops a
kernel-weighted coefficient recursion together with an online
kernel-bandwidth adaptation mechanism.

\section{Proposed KL-Adaptive DOA Estimator}
\label{sec:proposed_method}

This section develops the adaptive procedure for estimating the
spatial-filter coefficients and kernel bandwidth. A kernel-weighted
coefficient recursion is first introduced to reduce the influence of
large prediction errors under impulsive noise. The kernel bandwidth
is then adapted online from the prediction-error statistics, allowing
the degree of error attenuation to evolve during adaptation.

\subsection{Kernel-Weighted Spatial-Filter Adaptation}
\label{subsec:coefficient_adaptation}

The prediction error in \eqref{eq:prediction_error} drives the
adaptation of the spatial-filter coefficients. Under impulsive noise,
large residuals can induce disproportionately large coefficient
corrections in conventional error-driven adaptation. To limit their
influence, we introduce the Gaussian kernel weight
\begin{equation}
\kappa_h(e)
=
\exp\!\left(
-\frac{|e|^2}{2h^2}
\right),
\qquad h>0,
\label{eq:gaussian_kernel}
\end{equation}
where $h$ denotes the kernel bandwidth. The weight approaches unity
for small $|e|$ and decreases rapidly as the error magnitude
increases, thereby attenuating the contribution of large residuals.

Using \eqref{eq:gaussian_kernel}, the spatial-filter coefficients are
updated according to
\begin{equation}
\mathbf{w}_{n+1}
=
\mathbf{w}_n
+
\mu_w
\kappa_h\!\left(e(n)\right)
e^*(n)\bar{\mathbf{x}}_n,
\label{eq:kernel_weighted_update}
\end{equation}
where $\mu_w>0$ is the coefficient step size. Thus, the instantaneous
coefficient correction is scaled by
$\exp[-|e(n)|^2/(2h^2)]$, reducing the influence of large prediction
errors on the adaptation.

The bandwidth $h$ controls the resulting robustness--adaptation
tradeoff. A smaller bandwidth provides stronger attenuation of large
residuals, whereas a larger bandwidth weakens this attenuation and
makes \eqref{eq:kernel_weighted_update} approach a conventional
error-driven update. A fixed bandwidth may therefore be inadequate
as the prediction-error statistics evolve during adaptation,
motivating the online bandwidth adjustment developed next.

\subsection{KL-Based Kernel-Bandwidth Adaptation}
\label{subsec:bandwidth_adaptation}

The performance of the kernel-weighted recursion in
\eqref{eq:kernel_weighted_update} depends strongly on the kernel
bandwidth. Rather than prescribing this parameter a priori, we adapt
it from recent prediction errors using a KL
divergence criterion. Let $E\in\mathbb{C}$ denote the complex
prediction-error random variable with unknown density $f_E(e)$,
where $\mathbb{C}$ is identified with $\mathbb{R}^2$. For a
bandwidth $h>0$, let $\widehat{f}_h(e)$ denote a kernel estimate of
$f_E(e)$. The discrepancy between the two densities is measured by
\begin{equation}
D_{\mathrm{KL}}
\!\left(f_E\Vert\widehat{f}_h\right)
=
\int_{\mathbb{C}}
f_E(e)
\log
\frac{f_E(e)}
{\widehat{f}_h(e)}
\,d^2e ,
\label{eq:kl_divergence}
\end{equation}
where
$d^2e=d\Re\{e\}\,d\Im\{e\}$.
Since the term
$\int_{\mathbb{C}} f_E(e)\log f_E(e)\,d^2e$
is independent of $h$, minimizing
\eqref{eq:kl_divergence} with respect to $h$ is equivalent to
maximizing
\begin{equation}
\mathcal{J}(h)
=
\mathbb{E}_{E}
\left[
\log \widehat{f}_h(E)
\right].
\label{eq:kl_objective}
\end{equation}

At snapshot $n$, consider the most recent $L$ prediction errors,
$\mathcal{E}_n=\{e_{n-L+1},\ldots,e_n\}$. Because $f_E(e)$ is
unknown, the expectation in \eqref{eq:kl_objective} is approximated
from these samples. To exclude the self-contribution of each sample
from its own density estimate, we employ the leave-one-out kernel
density estimate
\begin{equation}
\widehat{f}_{-i,h}(e_i)
=
\frac{1}{L-1}
\sum_{\substack{j=1\\j\neq i}}^{L}
G_h(e_i-e_j),
\label{eq:loo_kde}
\end{equation}
where the isotropic Gaussian kernel on the complex plane is
\begin{equation}
G_h(z)
=
\frac{1}{2\pi h^2}
\exp\!\left(
-\frac{|z|^2}{2h^2}
\right),
\qquad z\in\mathbb{C}.
\label{eq:complex_gaussian_kernel}
\end{equation}
The normalization in \eqref{eq:complex_gaussian_kernel} is required
for density estimation on the two-dimensional complex plane; in
contrast, the coefficient recursion in
\eqref{eq:kernel_weighted_update} requires only the corresponding
exponential weighting term.

Using \eqref{eq:loo_kde}, a sample approximation of
\eqref{eq:kl_objective} is
\begin{equation}
\widehat{\mathcal{J}}_n(h)
=
\frac{1}{L}
\sum_{i=1}^{L}
\log
\left[
\frac{1}{L-1}
\sum_{\substack{j=1\\j\neq i}}^{L}
G_h(e_i-e_j)
\right].
\label{eq:sample_kl_criterion}
\end{equation}
Define the pairwise error difference as
$
\Delta e_{ij}=e_i-e_j
\label{eq:error_difference}
$
and the corresponding unnormalized Gaussian weight as
\begin{equation}
\gamma_{ij}(h)
=
\exp\!\left(
-\frac{|\Delta e_{ij}|^2}{2h^2}
\right).
\label{eq:pairwise_kernel}
\end{equation}
Differentiating \eqref{eq:sample_kl_criterion} with respect to the
real-valued bandwidth $h$ yields
\begin{equation}
\frac{\partial\widehat{\mathcal{J}}_n(h)}
{\partial h}
=
\frac{1}{L}
\sum_{i=1}^{L}
\frac{
\displaystyle
\sum_{\substack{j=1\\j\neq i}}^{L}
\gamma_{ij}(h)
\left(
\frac{|\Delta e_{ij}|^2}{h^3}
-\frac{2}{h}
\right)
}{
\displaystyle
\sum_{\substack{j=1\\j\neq i}}^{L}
\gamma_{ij}(h)
}.
\label{eq:bandwidth_gradient}
\end{equation}
Accordingly, the bandwidth is updated by gradient ascent on the
sample criterion:
\begin{equation}
h_{n+1}
=
\max\!\left\{
h_{\min},
\,
h_n
+
\mu_h
\left.
\frac{\partial\widehat{\mathcal{J}}_n(h)}
{\partial h}
\right|_{h=h_n}
\right\},
\label{eq:bandwidth_update}
\end{equation}
where $\mu_h>0$ denotes the bandwidth step size and $h_{\min}>0$
prevents the bandwidth from approaching zero.

The resulting recursion adapts the kernel bandwidth according to the
local distribution of recent complex prediction errors. The updated
bandwidth is subsequently used in the Gaussian weighting of the
spatial-filter recursion, allowing the degree of residual attenuation
to evolve during adaptation.

\subsection{Overall Adaptive DOA Procedure}
\label{subsec:overall_procedure}

The coefficient and bandwidth recursions are now combined into a
single online procedure. At snapshot $n$, the prediction error is
first computed using the current coefficient vector as
\begin{equation}
e(n)
=
x_0(n)-\mathbf{w}_n^H\bar{\mathbf{x}}_n .
\label{eq:online_prediction_error}
\end{equation}
Once at least $L$ prediction errors are available, the most recent
error window is used to evaluate
\eqref{eq:bandwidth_gradient}, and the bandwidth is updated according
to \eqref{eq:bandwidth_update}. During the initial acquisition period,
when fewer than $L$ errors are available, the prescribed initial
bandwidth $h_0$ is retained.

After the bandwidth update, define the instantaneous kernel weight as
\begin{equation}
\alpha_n
=
\exp\!\left(
-\frac{|e(n)|^2}{2h_{n+1}^2}
\right).
\label{eq:instantaneous_kernel_weight}
\end{equation}
The spatial-filter coefficients are then updated as
\begin{equation}
\mathbf{w}_{n+1}
=
\mathbf{w}_n
+
\mu_w
\alpha_n
e^*(n)\bar{\mathbf{x}}_n .
\label{eq:adaptive_coefficient_update}
\end{equation}
Thus, the prediction error generated by the current spatial filter
determines both the bandwidth adjustment and the subsequent
kernel-weighted coefficient correction.

For clarity, the complete procedure is summarized in
Algorithm~\ref{alg:proposed_method}. The bandwidth update requires
pairwise kernel evaluations within the $L$-sample error window,
whereas the coefficient update operates directly on the current
array snapshot.

\begin{algorithm}[!t]
\caption{Proposed KL-Adaptive DOA Estimator}
\label{alg:proposed_method}
\begin{algorithmic}[1]
\REQUIRE
$\{\mathbf{x}_n\}_{n=1}^{N}$,
$\mu_w>0$,
$\mu_h>0$,
$h_0>0$,
$h_{\min}>0$,
window length $L$
\STATE Initialize $\mathbf{w}_1=\mathbf{0}$ and $h_1=h_0$
\FOR{$n=1,\ldots,N$}
    \STATE Form
    $\bar{\mathbf{x}}_n=[x_1(n),\ldots,x_{M-1}(n)]^T$
    \STATE Compute
    $e(n)=x_0(n)-\mathbf{w}_n^H\bar{\mathbf{x}}_n$
    \IF{$n\geq L$}
        \STATE Form
        $\mathcal{E}_n=\{e(n-L+1),\ldots,e(n)\}$
        \STATE Evaluate
        $g_h(n)=
        \left.
        \partial\widehat{\mathcal{J}}_n(h)/\partial h
        \right|_{h=h_n}$
        using \eqref{eq:bandwidth_gradient}
        \STATE
        $h_{n+1}
        =
        \max\{h_{\min},\,h_n+\mu_h g_h(n)\}$
    \ELSE
        \STATE $h_{n+1}=h_n$
    \ENDIF
    \STATE
    $\alpha_n
    =
    \exp[-|e(n)|^2/(2h_{n+1}^2)]$
    \STATE
    $\mathbf{w}_{n+1}
    =
    \mathbf{w}_n+
    \mu_w\alpha_n e^*(n)\bar{\mathbf{x}}_n$
\ENDFOR
\STATE Construct
$P_N(\theta)
=
|1-\mathbf{w}_{N+1}^H\bar{\mathbf{a}}(\theta)|^{-2}$
over the prescribed angular search region
\STATE Estimate the DOAs from the dominant local maxima of
$P_N(\theta)$
\end{algorithmic}
\end{algorithm}

After the final snapshot, the adapted coefficient vector is used in
\eqref{eq:spatial_spectrum} to construct the spatial spectrum. When
the number of sources $Q$ is known, the $Q$ largest distinct local
maxima are selected as the DOA estimates.

\section{Performance Analysis}
\label{sec:analysis}

This section analyzes the coefficient behavior of the proposed
estimator. Because the kernel weight depends on the instantaneous
prediction error and the bandwidth is itself adaptive, an exact
stochastic analysis of the joint coefficient--bandwidth recursion is
generally intractable. We therefore characterize the coefficient
dynamics under standard adaptive-filtering approximations, treating
the bandwidth as locally quasi-stationary over the time scale of the
coefficient analysis. The resulting expressions provide conditions
for mean stability and, subsequently, an approximate characterization
of the mean-square behavior.

\subsection{Mean Behavior and Stability}
\label{subsec:mean_analysis}

Let $\mathbf{w}^{o}\in\mathbb{C}^{M-1}$ denote a deterministic
reference coefficient vector associated with the local
steady-state prediction relation, and define the coefficient-error
vector as
\begin{equation}
\widetilde{\mathbf{w}}_n
=
\mathbf{w}^{o}-\mathbf{w}_n .
\label{eq:weight_error}
\end{equation}
The residual corresponding to $\mathbf{w}^{o}$ is
\begin{equation}
d(n)
=
x_0(n)-\mathbf{w}^{oH}\bar{\mathbf{x}}_n .
\label{eq:reference_residual}
\end{equation}
Using \eqref{eq:prediction_error} and
\eqref{eq:weight_error}, the instantaneous prediction error can be
written as
\begin{equation}
e(n)
=
d(n)
+
\widetilde{\mathbf{w}}_n^H\bar{\mathbf{x}}_n .
\label{eq:error_decomposition}
\end{equation}

For the online recursion, define
\begin{equation}
\alpha_n
=
\exp\!\left(
-\frac{|e(n)|^2}{2h_{n+1}^2}
\right).
\label{eq:analysis_kernel_weight}
\end{equation}
Substituting \eqref{eq:error_decomposition} into the coefficient
update \eqref{eq:adaptive_coefficient_update} and using
$\widetilde{\mathbf{w}}_{n+1}
=\mathbf{w}^{o}-\mathbf{w}_{n+1}$ gives
\begin{equation}
\widetilde{\mathbf{w}}_{n+1}
=
\left(
\mathbf{I}_{M-1}
-
\mu_w\mathbf{R}_n
\right)
\widetilde{\mathbf{w}}_n
-
\mu_w\boldsymbol{\xi}_n ,
\label{eq:weight_error_recursion}
\end{equation}
where
$
\mathbf{R}_n
=
\alpha_n
\bar{\mathbf{x}}_n\bar{\mathbf{x}}_n^H
\label{eq:weighted_covariance_instantaneous}
$
and
$
\boldsymbol{\xi}_n
=
\alpha_n\bar{\mathbf{x}}_n d^*(n).
\label{eq:weighted_disturbance}
$

Equation \eqref{eq:weight_error_recursion} is an exact algebraic
representation of the coefficient recursion for a given
$\alpha_n$. Its statistical analysis is more involved because
$\alpha_n$ depends on $e(n)$ and is therefore generally correlated
with both $\widetilde{\mathbf{w}}_n$ and
$\bar{\mathbf{x}}_n$. To obtain a tractable mean model, we adopt the
usual local-independence approximation used in adaptive-filter
analysis and assume that, in the neighborhood of steady state, the
dependence between the current coefficient error and the weighted
instantaneous quantities can be neglected to first order. We further
assume that the reference vector $\mathbf{w}^{o}$ is chosen such that
the weighted residual satisfies the local orthogonality condition
\begin{equation}
\mathbb{E}
\left[
\alpha_n\bar{\mathbf{x}}_n d^*(n)
\right]
=
\mathbf{0}.
\label{eq:weighted_orthogonality}
\end{equation}

For a locally fixed bandwidth $h$, define the weighted correlation
matrix
\begin{equation}
\mathbf{G}(h)
=
\mathbb{E}
\left[
\alpha(h)
\bar{\mathbf{x}}\bar{\mathbf{x}}^H
\right].
\label{eq:weighted_correlation_matrix}
\end{equation}
Under the preceding approximations, taking expectations in
\eqref{eq:weight_error_recursion} yields
\begin{equation}
\mathbb{E}
\left[
\widetilde{\mathbf{w}}_{n+1}
\right]
\simeq
\left[
\mathbf{I}_{M-1}
-
\mu_w\mathbf{G}(h)
\right]
\mathbb{E}
\left[
\widetilde{\mathbf{w}}_n
\right].
\label{eq:mean_weight_error_recursion}
\end{equation}
Hence, for a fixed or locally quasi-stationary bandwidth, the
coefficient recursion is mean stable if
\begin{equation}
\rho
\left(
\mathbf{I}_{M-1}
-
\mu_w\mathbf{G}(h)
\right)
<1,
\label{eq:mean_stability_spectral}
\end{equation}
where $\rho(\cdot)$ denotes the spectral radius. If
$\mathbf{G}(h)$ is Hermitian positive definite, a sufficient and
necessary step-size condition for the approximate frozen-bandwidth
mean model is
\begin{equation}
0<\mu_w<
\frac{2}
{\lambda_{\max}\!\left(\mathbf{G}(h)\right)} .
\label{eq:mean_stability_bound}
\end{equation}

The condition in \eqref{eq:mean_stability_bound} characterizes the
local coefficient dynamics for a fixed or slowly varying bandwidth;
it should not be interpreted as a global stability guarantee for the
joint nonlinear recursion of $\mathbf{w}_n$ and $h_n$. For a
time-varying bandwidth, a stronger uniform contraction condition on
the sequence of mean transition matrices is required.

\subsection{Mean-Square Behavior}
\label{subsec:mean_square_analysis}

We next characterize the second-order behavior of the coefficient
error. Define
\begin{equation}
\mathbf{B}_n
=
\mathbf{I}_{M-1}
-
\mu_w\mathbf{R}_n .
\label{eq:transition_matrix}
\end{equation}
Then, \eqref{eq:weight_error_recursion} can be written compactly as
\begin{equation}
\widetilde{\mathbf{w}}_{n+1}
=
\mathbf{B}_n\widetilde{\mathbf{w}}_n
-
\mu_w\boldsymbol{\xi}_n .
\label{eq:compact_weight_error_recursion}
\end{equation}

For an arbitrary Hermitian positive-semidefinite weighting matrix
$\boldsymbol{\Sigma}$, define the weighted squared norm
$\|\mathbf{u}\|_{\boldsymbol{\Sigma}}^2
=
\mathbf{u}^H\boldsymbol{\Sigma}\mathbf{u}$.
Using \eqref{eq:compact_weight_error_recursion}, we obtain
\begin{align}
\|\widetilde{\mathbf{w}}_{n+1}\|_{\boldsymbol{\Sigma}}^2
={}&
\widetilde{\mathbf{w}}_n^H
\mathbf{B}_n^H\boldsymbol{\Sigma}\mathbf{B}_n
\widetilde{\mathbf{w}}_n
-
\mu_w
\widetilde{\mathbf{w}}_n^H
\mathbf{B}_n^H\boldsymbol{\Sigma}\boldsymbol{\xi}_n
\nonumber\\
&-
\mu_w
\boldsymbol{\xi}_n^H\boldsymbol{\Sigma}\mathbf{B}_n
\widetilde{\mathbf{w}}_n
+
\mu_w^2
\boldsymbol{\xi}_n^H\boldsymbol{\Sigma}\boldsymbol{\xi}_n .
\label{eq:weighted_energy_relation}
\end{align}

Under the same local-independence and quasi-stationarity
approximations adopted in the mean analysis, the two cross terms in
\eqref{eq:weighted_energy_relation} are neglected in expectation.
Consequently,
\begin{equation}
\mathbb{E}
\!\left[
\|\widetilde{\mathbf{w}}_{n+1}\|_{\boldsymbol{\Sigma}}^2
\right]
\simeq
\mathbb{E}
\!\left[
\|\widetilde{\mathbf{w}}_n\|_{\boldsymbol{\Sigma}'_n}^2
\right]
+
\mu_w^2
\operatorname{Tr}
\!\left(
\boldsymbol{\Sigma}\mathbf{Q}_n
\right),
\label{eq:mean_square_energy_relation}
\end{equation}
where
$
\boldsymbol{\Sigma}'_n
=
\mathbb{E}
\!\left[
\mathbf{B}_n^H
\boldsymbol{\Sigma}
\mathbf{B}_n
\right]
\label{eq:transformed_weighting_matrix}
$
and
$
\mathbf{Q}_n
=
\mathbb{E}
\!\left[
\boldsymbol{\xi}_n\boldsymbol{\xi}_n^H
\right].
\label{eq:disturbance_covariance}
$

To express \eqref{eq:mean_square_energy_relation} in vector form,
let
$\boldsymbol{\sigma}=\operatorname{vec}(\boldsymbol{\Sigma})$.
Using
$\operatorname{vec}(\mathbf{A}\mathbf{X}\mathbf{C})
=
(\mathbf{C}^{T}\otimes\mathbf{A})
\operatorname{vec}(\mathbf{X})$,
we obtain
$
\operatorname{vec}(\boldsymbol{\Sigma}'_n)
=
\mathbf{F}_n\boldsymbol{\sigma},
\label{eq:vectorized_weighting_recursion}
$
where
$
\mathbf{F}_n
=
\mathbb{E}
\left[
\mathbf{B}_n^{T}
\otimes
\mathbf{B}_n^{H}
\right].
\label{eq:mean_square_transition_matrix}
$
For a fixed or locally quasi-stationary bandwidth $h$, the
corresponding frozen second-order model is mean-square stable if
\begin{equation}
\rho\!\left(\mathbf{F}(h)\right)<1.
\label{eq:mean_square_stability}
\end{equation}
For a genuinely time-varying bandwidth, a sufficient uniform
stability condition is the existence of an induced matrix norm for
which
$
\sup_n \|\mathbf{F}_n\| < 1 .
\label{eq:uniform_mean_square_stability}
$

Under the frozen-bandwidth approximation, assume that
$\mathbf{F}_n\rightarrow\mathbf{F}_{\infty}$ and
$\mathbf{Q}_n\rightarrow\mathbf{Q}_{\infty}$, with
$\rho(\mathbf{F}_{\infty})<1$. At steady state,
\eqref{eq:mean_square_energy_relation} then gives
\begin{equation}
\mathbb{E}
\!\left[
\|\widetilde{\mathbf{w}}_{\infty}\|_{\boldsymbol{\Sigma}}^2
\right]
\simeq
\mu_w^2
\operatorname{vec}^{T}
\!\left(\mathbf{Q}_{\infty}^{T}\right)
\left(
\mathbf{I}_{(M-1)^2}-\mathbf{F}_{\infty}
\right)^{-1}
\operatorname{vec}(\boldsymbol{\Sigma}).
\label{eq:steady_state_weighted_mse}
\end{equation}
Setting
$\boldsymbol{\Sigma}=\mathbf{I}_{M-1}$ yields the approximate
steady-state MSD 
\begin{equation}
\begin{aligned}
\mathrm{MSD}_{\infty}
&\triangleq
\mathbb{E}\!\left[
\|\widetilde{\mathbf{w}}_{\infty}\|_2^2
\right]                                                     \\
&\simeq
\mu_w^2
\operatorname{vec}^{T}\!\left(\mathbf{Q}_{\infty}^{T}\right)
\left(
\mathbf{I}_{(M-1)^2}-\mathbf{F}_{\infty}
\right)^{-1}
\operatorname{vec}\!\left(\mathbf{I}_{M-1}\right).
\end{aligned}
\label{eq:steady_state_msd}
\end{equation}

Equations \eqref{eq:mean_square_stability} and
\eqref{eq:steady_state_msd} characterize the local second-order
behavior of the coefficient recursion under the stated
approximations. Because both $\mathbf{F}_{\infty}$ and
$\mathbf{Q}_{\infty}$ depend implicitly on the kernel weighting and
the limiting bandwidth statistics, these expressions characterize
the steady-state coefficient behavior without implying an exact
closed-form solution for the joint coefficient--bandwidth dynamics.

\subsection{Computational Complexity}
\label{subsec:complexity}

At each snapshot, computation of the prediction error and the
coefficient update in \eqref{eq:adaptive_coefficient_update} requires
operations that scale linearly with the number of array elements,
resulting in a complexity of $\mathcal{O}(M)$. Once $L$ prediction
errors are available, evaluation of the bandwidth gradient in
\eqref{eq:bandwidth_gradient} involves pairwise kernel computations
within the $L$-sample error window. A direct implementation therefore
requires $\mathcal{O}(L^2)$ operations per bandwidth update.
Consequently, the adaptive processing cost over $N$ snapshots is
$
\mathcal{O}\!\left(
N\left(M+L^2\right)
\right).
\label{eq:adaptive_complexity}
$
After adaptation, evaluating the spatial spectrum in
\eqref{eq:spatial_spectrum} over $P$ candidate angles requires
$\mathcal{O}(PM)$ operations. Hence, the overall complexity is
$
\mathcal{O}\!\left(
N\left(M+L^2\right)+PM
\right).
\label{eq:total_complexity}
$
The additional computational cost relative to fixed-bandwidth
coefficient adaptation is primarily due to the pairwise operations
required for bandwidth adaptation. Since these operations are
performed over a finite error window, the associated cost is
controlled by the window length $L$.

\section{Numerical Results}
\label{sec:results}

This section evaluates the proposed estimator under impulsive-noise
conditions. Unless otherwise stated, the simulation settings are
those summarized in Table~\ref{tab:simulation_parameters}.

\subsection{Simulation Setup}
\label{subsec:simulation_setup}

A ULA with half-wavelength interelement spacing, $d=\lambda/2$, is
considered. The source signals are mutually independent zero-mean
circular complex Gaussian sequences with unit variance. The noise at
each sensor follows the Gaussian-mixture model
\begin{equation}
v_m(n)
\sim
(1-\varepsilon)\mathcal{CN}(0,\sigma_v^2)
+
\varepsilon\mathcal{CN}(0,\kappa\sigma_v^2),
\label{eq:impulsive_noise_model}
\end{equation}
where $\mathcal{CN}(0,\sigma^2)$ denotes a zero-mean circularly
symmetric complex Gaussian distribution with variance $\sigma^2$,
$\epsilon$ is the impulse probability, and $\kappa>1$ controls the
impulsive component. The corresponding total noise
variance is
\begin{equation}
\sigma_{\mathrm{mix}}^2
=
\big[(1-\varepsilon)+\varepsilon\kappa\big]\sigma_v^2.
\label{eq:mixture_noise_variance}
\end{equation}
Accordingly, the signal-to-noise ratio (SNR) is defined with respect
to the total mixture noise power as
\begin{equation}
\mathrm{SNR}
=
10\log_{10}
\frac{\mathbb{E}[|s_q(n)|^2]}
{\sigma_{\mathrm{mix}}^2}
\quad \mathrm{dB}.
\label{eq:snr_definition}
\end{equation}

\begin{table}[!t]
\centering
\caption{Default Simulation Parameters}
\label{tab:simulation_parameters}
\renewcommand{\arraystretch}{1.10}
\setlength{\tabcolsep}{5pt}
\begin{tabular}{lc}
\hline
\textbf{Parameter} & \textbf{Value} \\
\hline
Impulse probability, $\varepsilon$ & $0.1$ \\
Impulse power factor, $\kappa$ & $30$ \\
Initial bandwidth, $h_0$ & $3$ \\
Coefficient step size, $\mu_w$ & $4\times10^{-3}$ \\
Bandwidth step size, $\mu_h$ & $0.1$ \\
Error-window length, $L$ & $10$ \\
Number of snapshots, $N$ & $1000$ \\
Monte Carlo trials, $N_{\mathrm{MC}}$ & $1000$ \\
\hline
\end{tabular}
\end{table}

DOA accuracy is measured by the root-mean-square error (RMSE) over
$N_{\rm MC}$ independent trials,
\begin{equation}
\mathrm{RMSE}
=
\sqrt{
\frac{1}{Q N_{\mathrm{MC}}}
\sum_{r=1}^{N_{\mathrm{MC}}}
\sum_{q=1}^{Q}
\left(
\hat{\theta}_{q,r}-\theta_q
\right)^2
},
\label{eq:rmse_definition}
\end{equation}
where the estimated DOAs are matched to the true DOAs before the
error is evaluated. Parameters varied in individual experiments are
specified with the corresponding results.

\subsection{Adaptive Bandwidth and Spatial-Spectrum Behavior}
\label{subsec:adaptation_spectrum}

We first examine the adaptation behavior of the proposed estimator.
Fig.~2(a) compares the coefficient MSD obtained with fixed and
adaptive kernel bandwidths, while Fig.~2(b) shows the corresponding
bandwidth evolution. The adaptive bandwidth departs from its initial
value during the transient period and subsequently settles within a
stable range. The accompanying MSD behavior indicates improved
steady-state coefficient accuracy relative to the fixed-bandwidth
setting.

\begin{figure}[!t]
    \centering
    \subfloat[]{
        \includegraphics[width=8.3cm,height=3cm]
        {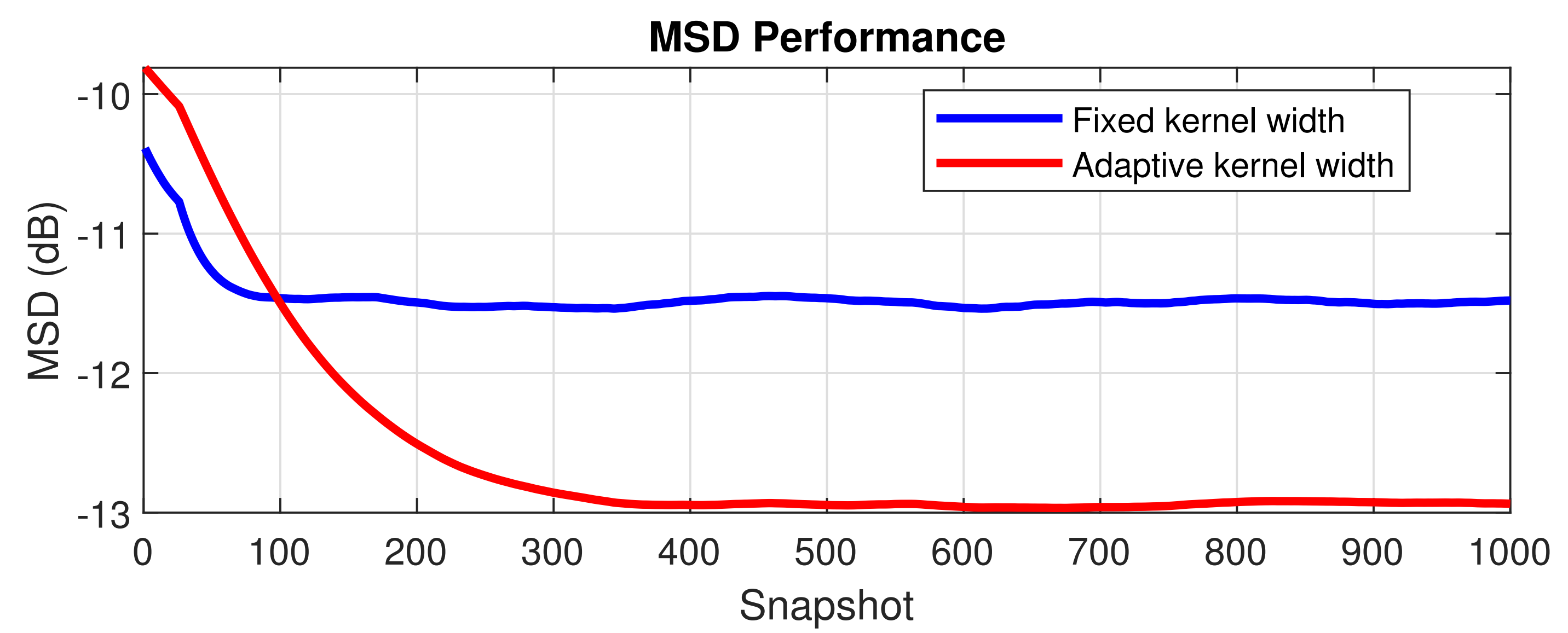}
        \label{fig:bandwidth_msd}}

    \vspace{-1mm}

    \subfloat[]{
        \includegraphics[width=8.0cm,height=3cm]
        {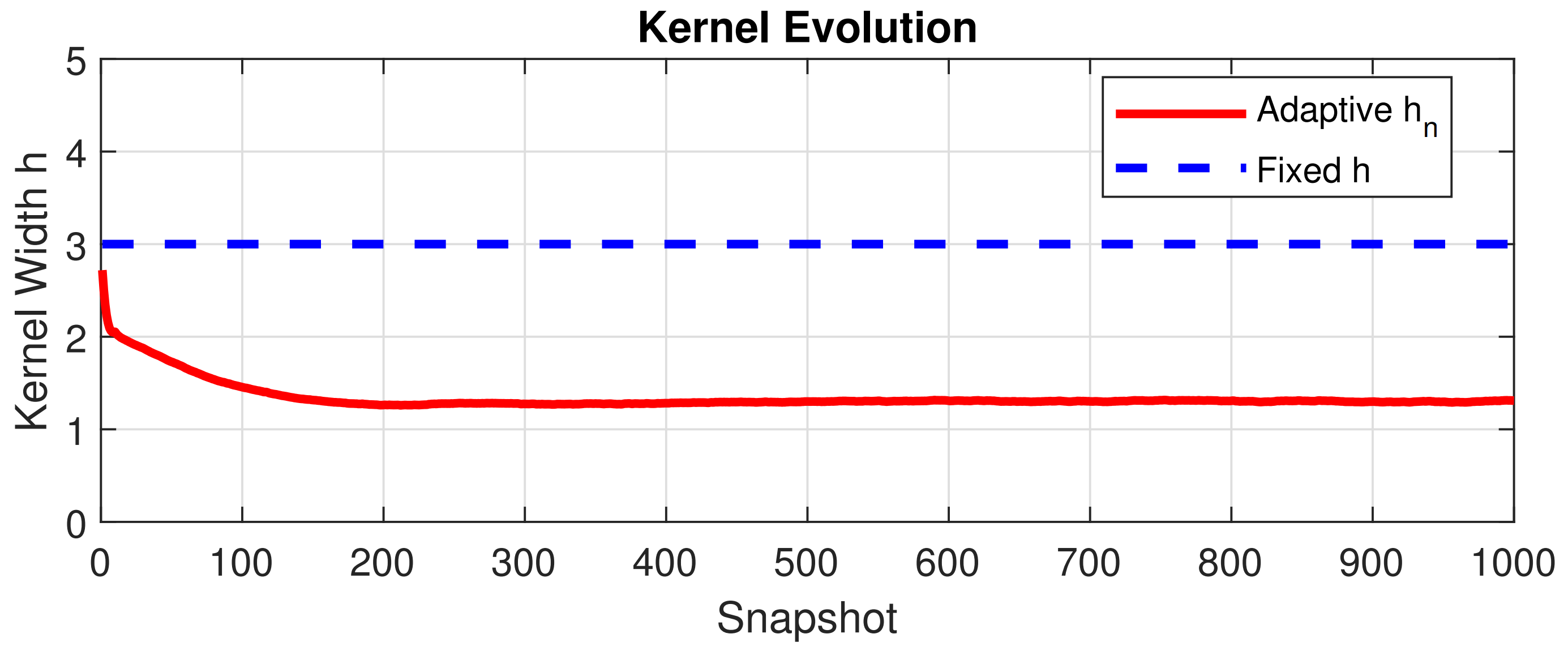}
        \label{fig:bandwidth_evolution}}

    \caption{Adaptation behavior of the proposed estimator:
    (a) coefficient MSD and (b) kernel-bandwidth evolution.}
    \label{fig:bandwidth_adaptation}
\end{figure}

\begin{figure}[!t]
    \centering

    \subfloat[]{
        \includegraphics[width=8.8cm,height=4.8cm]
        {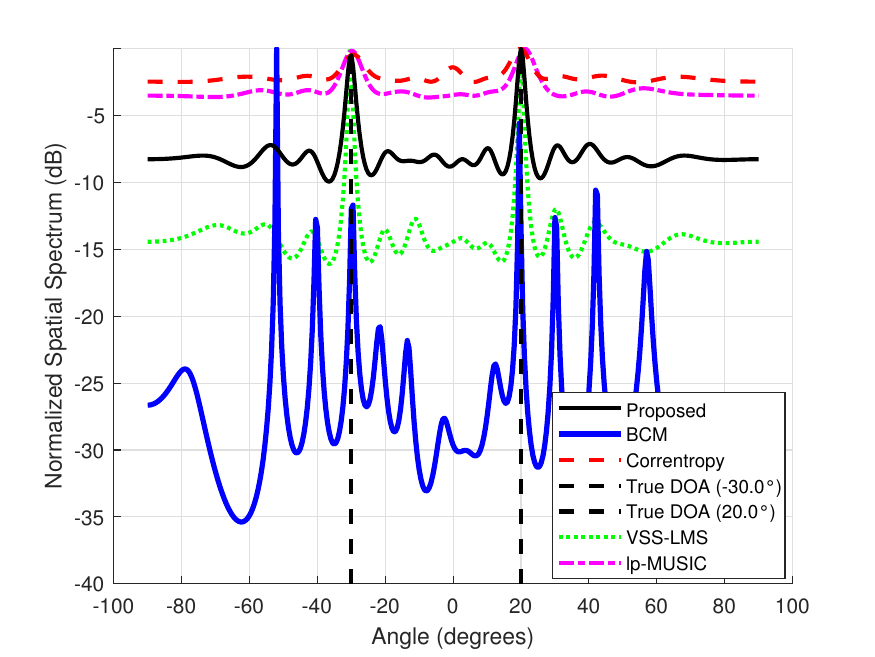}
        \label{fig:spectrum_low_snr}}

    \vspace{-1mm}

    \subfloat[]{
        \includegraphics[width=8.8cm,height=4.8cm]
        {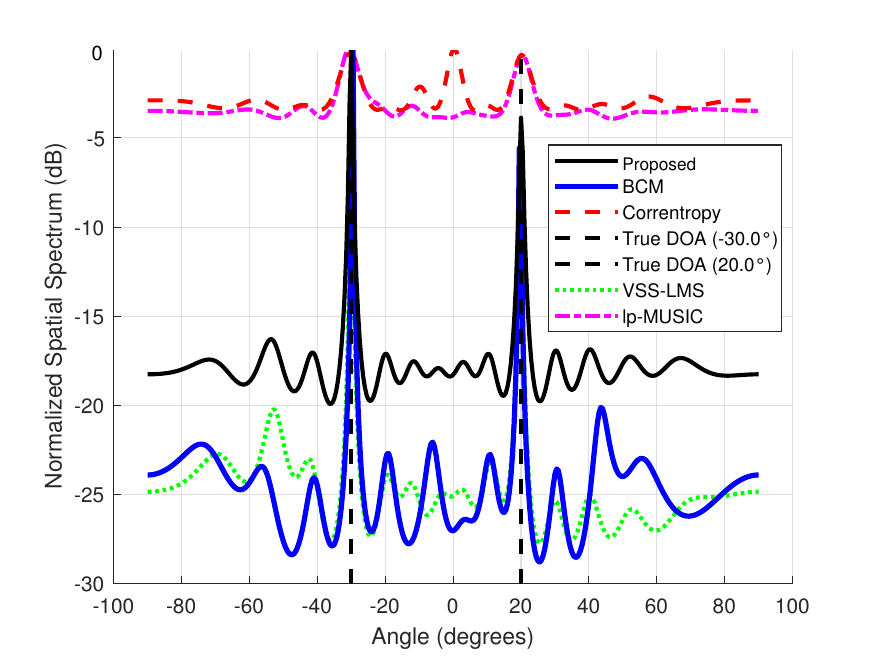}
        \label{fig:spectrum_high_snr}}

   \caption{Spatial spectra under impulsive noise at
(a) SNR $=4$ dB and (b) SNR $=10$ dB.}
    \label{fig:spatial_spectra}
\end{figure}

The corresponding spatial spectra are shown in
Fig.~\ref{fig:spatial_spectra} for two representative SNR levels.
At the lower SNR, the impulsive disturbance produces a more
challenging estimation condition, whereas increasing the SNR yields
sharper and more clearly separated spectral peaks. In both cases, the
dominant peaks produced by the proposed estimator remain aligned with
the true source directions, demonstrating that the adapted
spatial-filter coefficients preserve the directional information
required for DOA recovery.

\subsection{DOA Estimation Performance}
\label{subsec:doa_performance}

\begin{figure*}[!t]
    \centering
    \subfloat[]{
        \includegraphics[width=9.1cm,height=6.1cm]
        {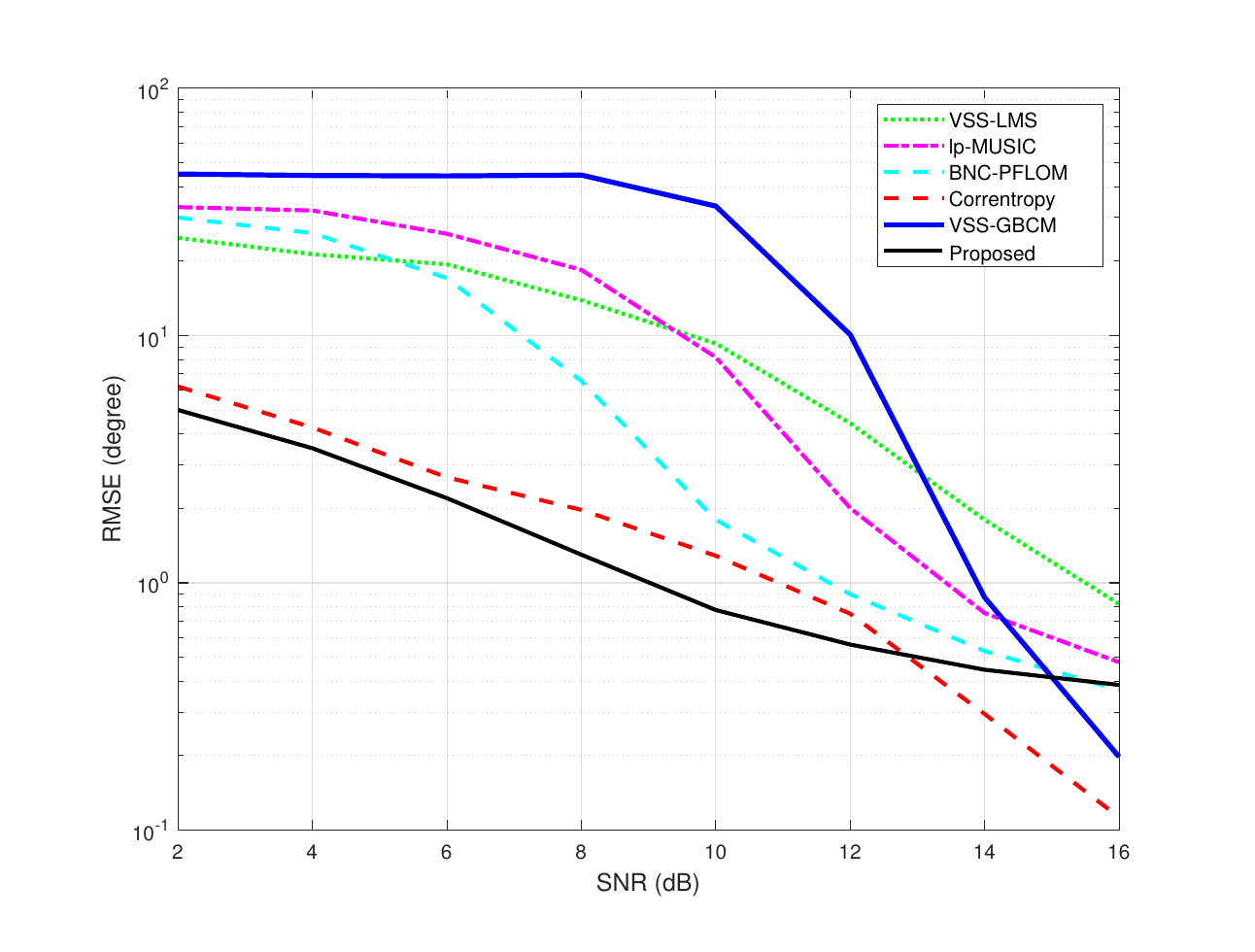}
        \label{fig:rmse_snr}}
    \hfill
    \subfloat[]{
        \includegraphics[width=9cm,height=5.6cm]
        {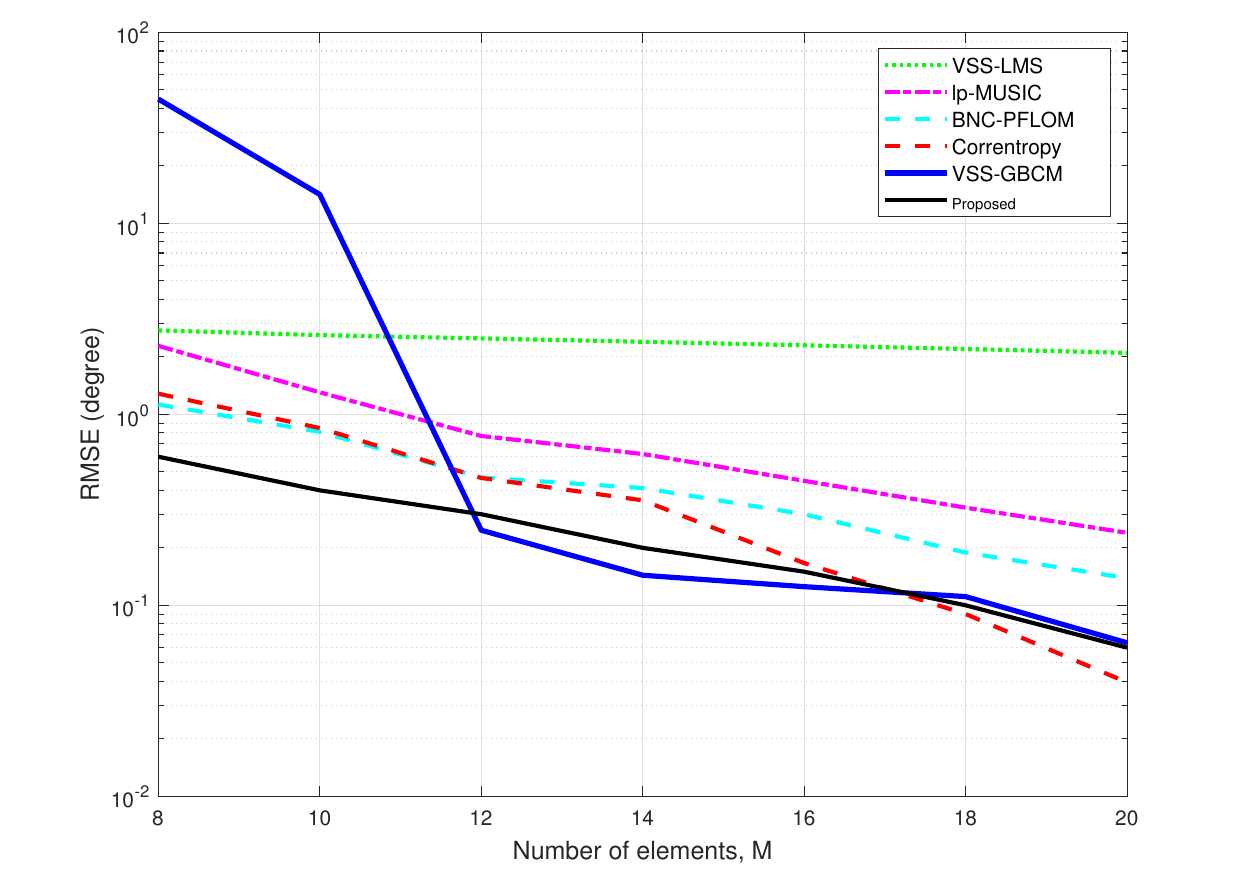}
        \label{fig:rmse_elements}}

    \subfloat[]{
        \includegraphics[width=9cm,height=5.6cm]
        {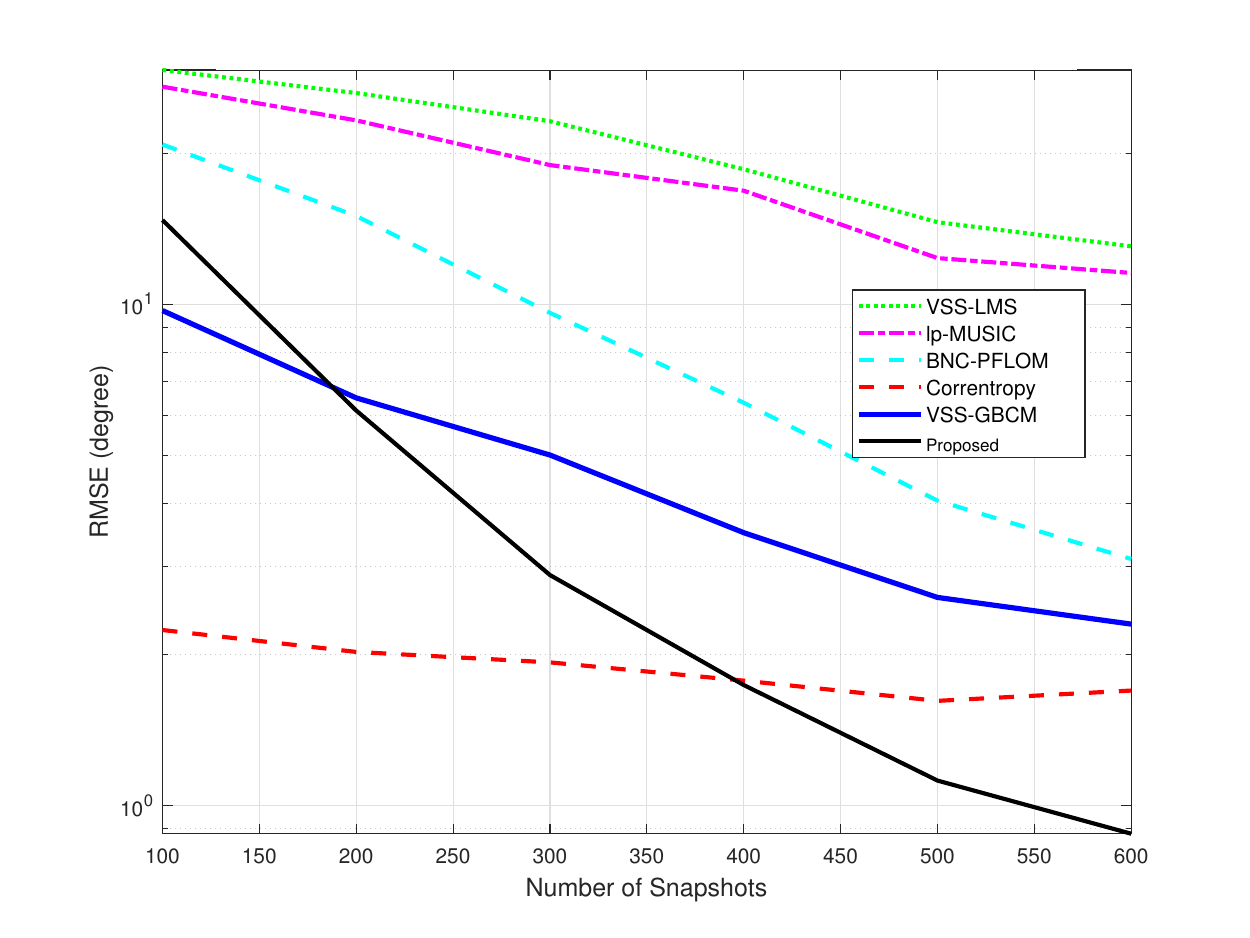}
        \label{fig:rmse_snapshots}}
    \hfill
    \subfloat[]{
        \includegraphics[width=9cm,height=5.6cm]
        {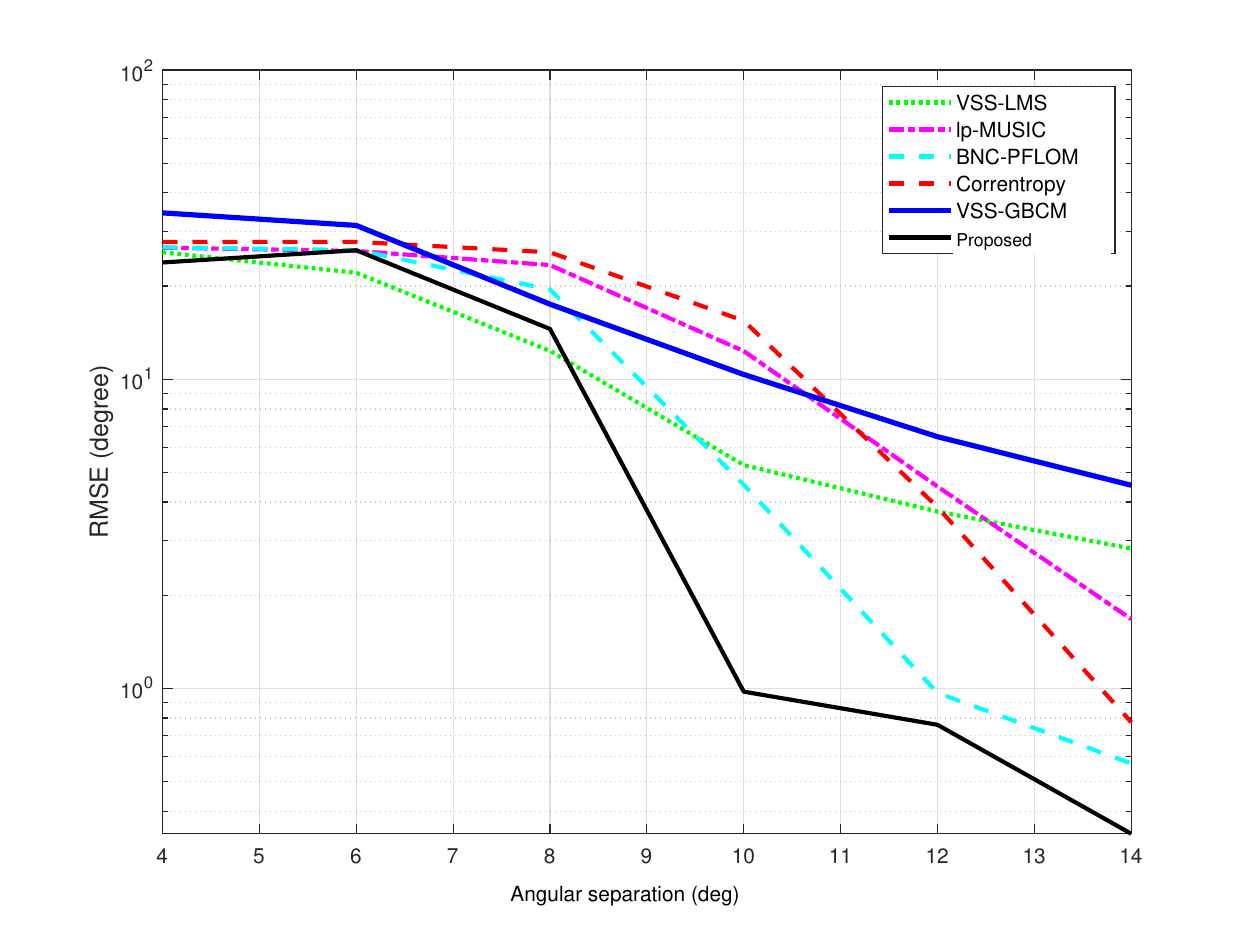}
        \label{fig:rmse_separation}}

    \caption{DOA estimation RMSE versus
    (a) SNR, (b) number of array elements,
    (c) number of snapshots, and
    (d) angular separation.}
    \label{fig:rmse_performance}
\end{figure*}

We next evaluate the DOA estimation accuracy under different
operating conditions. Fig.~\ref{fig:rmse_performance}(a) shows the
RMSE versus SNR. The proposed estimator provides its most pronounced
improvement in the low-to-moderate-SNR regime, where impulsive
observations have a stronger effect on the estimation process. As the
SNR increases, the performance gap narrows and the competing
kernel-based methods become comparable, indicating that the main
benefit of adaptive bandwidth control arises under more adverse noise
conditions.
Fig.~\ref{fig:rmse_performance}(b) examines the effect of the number
of array elements. Increasing $M$ generally improves the estimation
accuracy by providing greater spatial information. The proposed
estimator maintains competitive performance across the considered
array sizes, with larger gains under the more challenging
configurations.
The effect of the number of snapshots is shown in
Fig.~\ref{fig:rmse_performance}(c). The RMSE decreases as more
observations become available, reflecting the improved statistical
information available to the adaptive estimator. The proposed method
retains its advantage over most competing approaches over the
considered snapshot range.
Finally, Fig.~\ref{fig:rmse_performance}(d) evaluates angular
resolution by varying the separation between two sources. As the
sources become more closely spaced, the estimation problem becomes
increasingly difficult and the RMSE increases. The proposed estimator
exhibits improved robustness in this regime, while the differences
among the methods diminish as the angular separation increases.

\subsection{Parameter Sensitivity}
\label{subsec:parameter_sensitivity}

\begin{figure*}[!t]
    \centering

    \subfloat[]{
        \includegraphics[width=4.5cm,height=4.0cm]
        {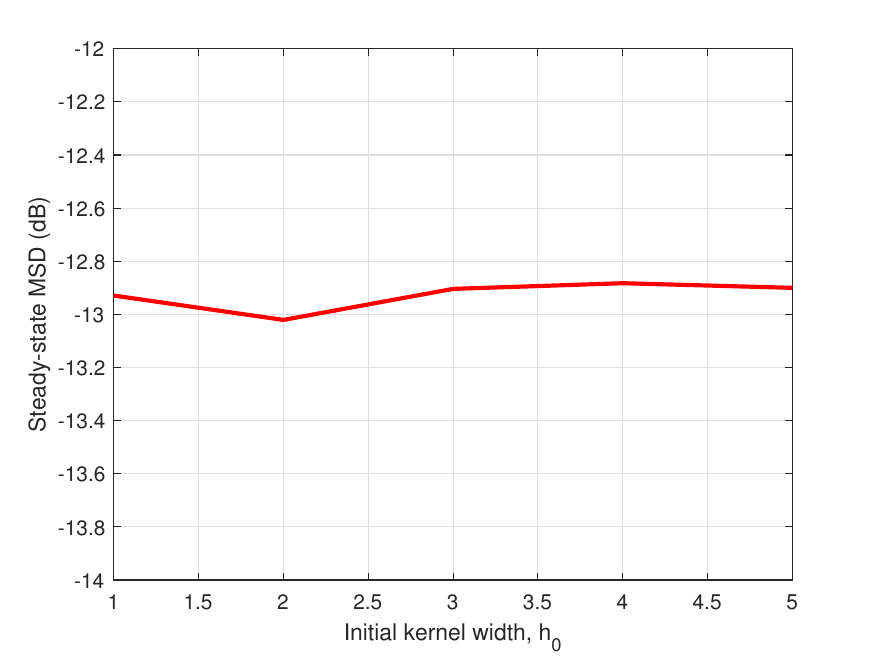}
        \label{fig:sensitivity_h0}}
    \hfill
    \subfloat[]{
        \includegraphics[width=4.5cm,height=4.0cm]
        {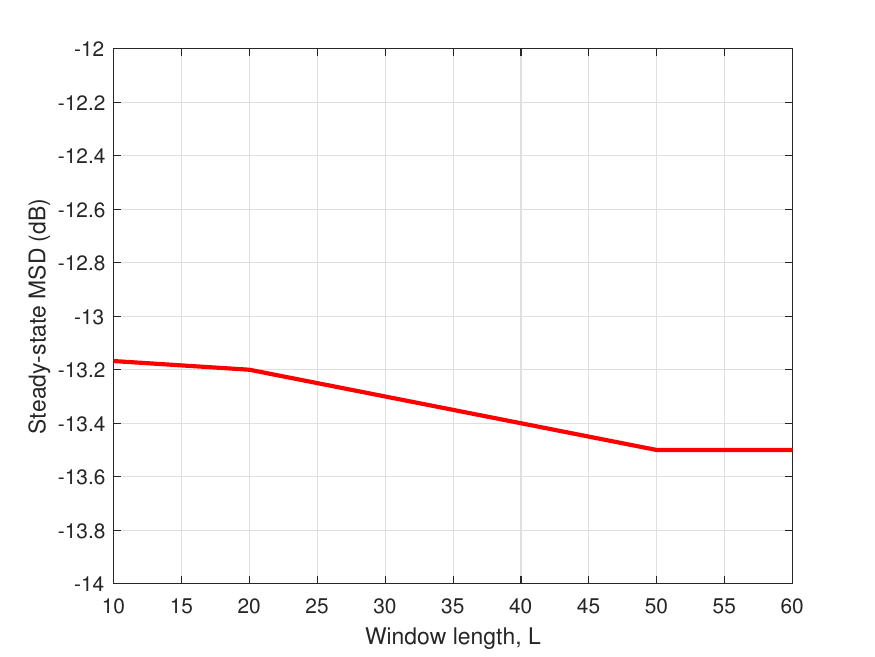}
        \label{fig:sensitivity_L}}
    \hfill
    \subfloat[]{
        \includegraphics[width=4.5cm,height=4.0cm]
        {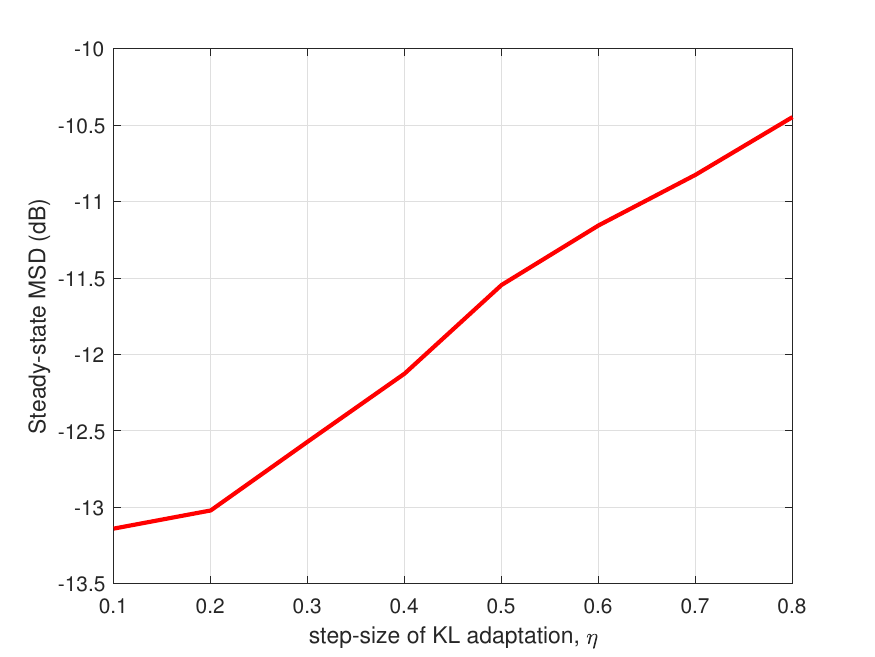}
        \label{fig:sensitivity_muh}}
    \hfill
    \subfloat[]{
        \includegraphics[width=4.5cm,height=4.0cm]
        {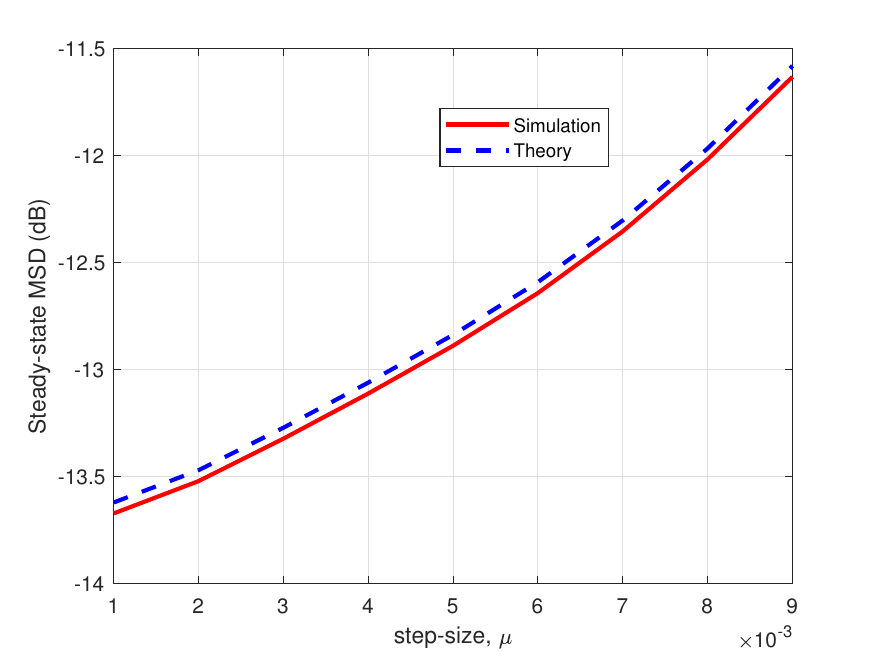}
        \label{fig:sensitivity_muw}}

    \caption{Sensitivity of the proposed estimator to
    (a) the initial bandwidth $h_0$,
    (b) the error-window length $L$,
    (c) the bandwidth step size $\mu_h$, and
    (d) the coefficient step size $\mu_w$.}

    \label{fig:parameter_sensitivity}
\end{figure*}

We next examine the sensitivity of the proposed estimator to its
principal adaptation parameters. Fig.~\ref{fig:parameter_sensitivity}
(a) shows the effect of the initial bandwidth $h_0$.
The relatively small variation in the steady-state MSD over the
considered range indicates reduced sensitivity to the initial
kernel bandwidth. The effects of the
error-window length $L$ and bandwidth step size $\mu_h$ are shown in
Fig.~\ref{fig:parameter_sensitivity}(b) and (c), respectively. These
parameters govern the responsiveness of the bandwidth recursion:
the window length controls the amount of recent error information
used in the KL criterion, while $\mu_h$ controls the magnitude of
each bandwidth correction.
Fig.~\ref{fig:parameter_sensitivity}(d) illustrates the effect of the
coefficient step size $\mu_w$. Small values lead to slower coefficient
adaptation, whereas excessively large values increase the
steady-state estimation error and can degrade the stability of the
adaptive recursion. Overall, the results indicate that satisfactory
performance is maintained over a practical range of parameter
settings, while the best operating point reflects the expected
tradeoff between adaptation speed and steady-state accuracy.

\subsection{Mean-Square Convergence Behavior}
\label{subsec:msd_behavior}

\begin{figure}[!t]
    \centering
\includegraphics[width=9cm,height=4.85cm]
    {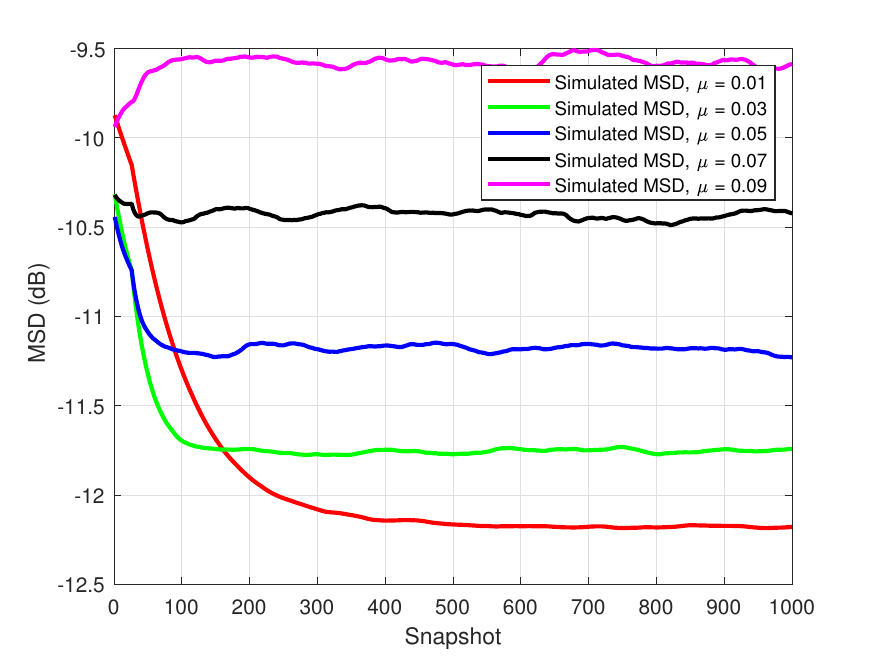}
    \caption{MSD convergence for different coefficient step sizes
    $\mu_w$.}
    \label{fig:msd_convergence}
\end{figure}

Finally, we examine the coefficient mean-square behavior to relate
the numerical results to the analysis in Section~\ref{sec:analysis}.
Fig.~\ref{fig:msd_convergence} shows the MSD trajectories for
different coefficient step sizes $\mu_w$. Smaller step sizes yield
slower convergence but lower steady-state MSD, whereas increasing
$\mu_w$ accelerates the transient adaptation at the expense of a
higher steady-state error. This behavior is consistent with the
speed--accuracy tradeoff predicted by the coefficient-error
recursion and the steady-state characterization in
\eqref{eq:steady_state_msd}.
The results also illustrate the local nature of the stability
analysis. Over the tested step-size range, the coefficient recursion
remains empirically stable, although its transient and steady-state
behavior varies substantially with $\mu_w$. These observations are
consistent with the approximate mean and mean-square conditions in
\eqref{eq:mean_stability_bound} and
\eqref{eq:mean_square_stability}, without implying a global stability
guarantee for the joint coefficient--bandwidth recursion.

\section{Conclusion}
\label{sec:conclusion}

This paper developed an adaptive DOA estimator for impulsive-noise
environments by combining kernel-weighted spatial-filter adaptation
with KL-based online kernel-bandwidth adjustment. A sample-based KL
criterion was formulated using a complex-valued kernel density
estimate, leading to a recursive bandwidth update driven by recent
prediction errors. Mean and mean-square analyses were developed under
standard local-independence and quasi-stationarity approximations to
characterize the coefficient convergence and steady-state behavior.
Numerical results under Gaussian-mixture impulsive noise showed that
the proposed estimator provides improved DOA accuracy in challenging
operating conditions, with the most pronounced gains at
low-to-moderate SNRs, while reducing sensitivity to the initial
kernel bandwidth. The results also demonstrated the expected
tradeoff between adaptation speed and steady-state accuracy. The
analysis characterizes the local coefficient dynamics rather than
providing a global stability guarantee for the joint
coefficient--bandwidth recursion, leaving a more complete joint
stochastic analysis as a direction for future work.

	\bibliographystyle{IEEEtran}
	\bibliography{Main_Document}

	\vfill
	
\end{document}